%% file: btVW.tex
\documentstyle[preprint,seceq]{jpsj}

\input{epsfile.inc}

\def\runtitle{Superfluid-Insulator Transition
	      of Interacting Multi-Component Bosons}
\def\runauthor{Yukitoshi {\sc Motome} and Masatoshi {\sc Imada}}

\def\sf{\rm}

\title{
Superfluid-Insulator Transition \\
of Interacting Multi-Component Bosons \\
\---- Gutzwiller Variational and 
      Quantum Monte Carlo Study \----}
\author{
Yukitoshi {\sc Motome} and Masatoshi {\sc Imada}
}
\inst{
Institute for Solid State Physics, University of Tokyo, \\
Roppongi 7-22-1, Minato-ku, Tokyo 106
}
\recdate{
February 14, 1996
}

\abst{
Various types of superfluid-insulator transitions are investigated
for two-component lattice boson systems in two dimensions
with on-site hard-core repulsion and
the component-dependent intersite interaction.
The mean-field phase diagram is obtained 
by the Gutzwiller-type variational technique
in the plane of filling and interaction parameters.
Various ground-state properties are also studied
by the quantum Monte Carlo method.
Our model exhibits two types of diagonal long-range orders;
the density order around the density $n=1/2$ and 
the Ising-type component order near $n=1$.
The quantum Monte Carlo results for
the transitions from the superfluid state to these two ordered states 
show marked contrast with the Gutzwiller results.
Namely, although they are both accompanied by phase separation
into commensurate ($n=1/2$ or $n=1$) and incommensurate density phases,
these transitions are both continuous.
The continuous growth of the component correlation
severely suppresses the superfluidity
as well as the inverse of the effective mass
in the critical region of the component order transition
in contrast to the persistence of the superfluidity
in the density-ordered state.
We propose a mechanism of the mass enhancement observed
even far from the Mott insulating filling $n=1$,
when the Ising-type component order persists into $n \neq 1$.
Possible relevance of this type of mass enhancement in other systems
is also discussed.
}

\kword{superfluid-insulator transition, 
       Mott transition,
       two-component boson system,
       component long-range order,
       Ising exchange,
       phase separation,
       strong mass enhancement}

\begin{document}
\sloppy
\maketitle

\input{intro.tex}

\input{model.tex}

\input{method.tex}

\input{result.tex}

\input{discussion.tex}

\input{summary.tex}

\input{acknowledge.tex}

\end{document}

%% file: epsfile.inc
\catcode`\@=11
\newcount\@arga
\newcount\@argb
\newcount\@argc
\newcount\@ctmpa
\newcount\@ctmpb
\newcount\@ctmpc
\newcount\@ctmpd
\newcount\@ctmpe
\newdimen\@darg
\newdimen\@bblen
\newif\if@bbllx
\newif\if@bblly
\newif\if@bburx
\newif\if@bbury
\newif\if@height
\newif\if@width
\newif\if@scale
\newif\ifno@bb
\newif\ifepsfdraft
\epsfdraftfalse
\def\@setpsfile#1{
                \typeout{epsf:[#1]}
                \def\@psfile{#1}
}
\def\@setpsheight#1{
                \@heighttrue
                \@darg=#1\relax
                \edef\@psheight{\number\@darg}
}
\def\@setpswidth#1{
                \@widthtrue
                \@darg=#1\relax
                \edef\@pswidth{\number\@darg}
}
\def\@setpsscale#1{
                \@scaletrue
                \def\@pshscale{#1}
                \def\@psvscale{#1}
                \@bblen=#1pt\relax
                \@bblen=1000\@bblen
                \def\@texhscale{\expandafter\remove@dim\the\@bblen}
                \let\@texvscale=\@texhscale
}
\def\@setpshscale#1{
                \@scaletrue
                \def\@pshscale{#1}
                \@bblen=#1pt\relax
                \@bblen=1000\@bblen
                \def\@texhscale{\expandafter\remove@dim\the\@bblen}
}

\def\@setpsvscale#1{
                \@scaletrue
                \def\@psvscale{#1}
                \@bblen=#1pt\relax
                \@bblen=1000\@bblen
                \def\@texvscale{\expandafter\remove@dim\the\@bblen}
}

\def\@setparms#1=#2,{\@nameuse{@setps#1}{#2}}
\def\ps@init@parms{
                \@heightfalse \@widthfalse
                \no@bbfalse
                \def\@psbbllx{}\def\@psbblly{}
                \def\@psbburx{}\def\@psbbury{}
                \def\@psheight{}\def\@pswidth{}
                \def\@pshscale{1}\def\@psvscale{1}
                \def\@texhscale{1000}\def\@texvscale{1000}
                \def\@psfile{}
                \def\@sc{}
}
\def\parse@ps@parms#1{
                \@for\@epsfile:=#1\do
                   {\expandafter\@setparms\@epsfile,}}
\newif\ifnot@eof
\newread\ps@stream
\def\bb@search{
        \openin\ps@stream=\@psfile
        \no@bbtrue
        \not@eoftrue
        \catcode`\%=12\relax
        \ifeof\ps@stream\typeout{epsf: File not found}\fi
        \loop
                \read\ps@stream to \line@in
                \global\toks200=\expandafter{\line@in}\relax
                \ifeof\ps@stream \not@eoffalse \fi
                \@bbtest{\toks200}\relax
                \if@bbmatch\not@eoffalse\expandafter\bb@cull\the\toks200\fi
        \ifnot@eof \repeat
        \catcode`\%=14
}       

\catcode`\%=12
\newif\if@bbmatch
\def\@bbtest#1{\expandafter\@a@\the#1
\long\def\@a@#1
        \ifx\@bbtest#2\@bbmatchfalse\else\@bbmatchtrue\fi}
\def\bb@cull 
        \@ifnextchar\space{\@latexbug}{\bb@extract}}
\def\bb@extract #1 #2 #3 #4 {
        \message{BoundingBox: (#1bp,#2bp)--(#3bp,#4bp)}
        \@darg=#1 bp\edef\@psbbllx{\number\@darg}
        \@darg=#2 bp\edef\@psbblly{\number\@darg}
        \@darg=#3 bp\edef\@psbburx{\number\@darg}
        \@darg=#4 bp\edef\@psbbury{\number\@darg}
        \no@bbfalse
}
\catcode`\%=14

\def\compute@bb{
                \bb@search
                \ifno@bb \typeout{epsf: No BoundingBox}
                \stop
                \else
                \@arga=\@psbburx
                \advance\@arga by -\@psbbllx
                \edef\@bbw{\number\@arga}
                \@arga=\@psbbury
                \advance\@arga by -\@psbblly
                \edef\@bbh{\number\@arga}
                \fi
}
\def\in@hundreds#1#2#3{\@argb=#2 \@argc=#3
                     \@ctmpa=\@argb     
                     \divide\@ctmpa by \@argc
                     \@ctmpb=\@ctmpa
                     \multiply\@ctmpb by \@argc
                     \advance\@argb by -\@ctmpb
                     \multiply\@argb by 10
                     \@ctmpb=\@argb     
                     \divide\@ctmpb by \@argc
                     \@ctmpc=\@ctmpb
                     \multiply\@ctmpc by \@argc
                     \advance\@argb by -\@ctmpc
                     \multiply\@argb by 10
                     \@ctmpc=\@argb     
                     \divide\@ctmpc by \@argc
                     \@arga=#1\@ctmpe=0
                     \@ctmpd=\@arga
                        \multiply\@ctmpd by \@ctmpa
                        \advance\@ctmpe by \@ctmpd
                     \@ctmpd=\@arga
                        \divide\@ctmpd by 10
                        \multiply\@ctmpd by \@ctmpb
                        \advance\@ctmpe by \@ctmpd
                     \@ctmpd=\@arga
                        \divide\@ctmpd by 100
                        \multiply\@ctmpd by \@ctmpc
                        \advance\@ctmpe by \@ctmpd
                     \edef\@result{\number\@ctmpe}
}
\def\compute@wfromh{
                \in@hundreds{\@psheight}{\@bbw}{\@bbh}
                \edef\@pswidth{\@result}
}
\def\compute@hfromw{
                \in@hundreds{\@pswidth}{\@bbh}{\@bbw}
                \edef\@psheight{\@result}
}
\def\compute@handw{
        \if@height 
                \if@width
                \else
                        \compute@wfromh
                \fi
        \else 
                \if@width
                        \compute@hfromw
                \else
                        \if@scale
                                \in@hundreds{\@texvscale}{\@bbh}{1000}
                                \let\@bbh=\@result
                                \in@hundreds{\@texhscale}{\@bbw}{1000}
                                \let\@bbw=\@result
                        \fi
                                \edef\@psheight{\@bbh}
                                \edef\@pswidth{\@bbw}
                \fi
        \fi
}
{\catcode`\p=12\catcode`\t=12
\gdef\remove@dim#1.#2pt{#1}}

\def\compute@sizes{
        \compute@bb
        \compute@handw
}
\def\epsfile#1{
        \ps@init@parms
        \parse@ps@parms{#1}
        \compute@sizes
        \@arga=\@psheight
        \divide\@arga by 65536
        \edef\@psvsize{\number\@arga}
        \@arga=\@pswidth
        \divide\@arga by 65536
        \edef\@pshsize{\number\@arga}
        \message{=>(\@pshsize bp,\@psvsize bp)}
        \leavevmode
        \vbox to \@psheight true sp{
                \hbox to \@pswidth true sp{
                \ifepsfdraft\hss\@psfile\hss\else
                \if@height 
                        \if@width
                                \special{epsfile=\@psfile \space 
                                hsize=\@pshsize \space
                                vsize=\@psvsize \space}
                        \else
                                \special{epsfile=\@psfile \space 
                                vsize=\@psvsize \space}
                        \fi
                \else 
                        \if@width
                                \special{epsfile=\@psfile \space 
                                hsize=\@pshsize \space}
                        \else
                                \if@scale
                                        \special{epsfile=\@psfile \space
                                        vscale=\@psvscale \space
                                        hscale=\@pshscale \space}
                                \else
                                        \special{epsfile=\@psfile \space}
                                \fi
                        \fi
                \fi
                \hfil\fi
                }
        \vfil
        }
}

\catcode`\@=12

%% file: intro.tex
\section{Introduction}

Recently, the scaling theory of the metal-insulator transition
has been proposed \cite{Imada1},
where the critical nature of the transition depends on
characters of both of the insulating and the metallic states.
An origin of the difference in characters is due to a variety of
the spin degrees of freedom such as
the antiferromagnetic long-range order or the spin excitation gap.
These differences cause 
various different types of metal-insulator transitions.
This scaling theory has been combined 
with numerical results of the two-dimensional Hubbard model
leading to the finding of a new universality class
in the Mott transitions.
Numerical calculations show following singular properties:
(1) In the metallic state, the charge susceptibility $\chi_{c}$
is inversely proportional to the hole density $\delta=1-n$
near the Mott transition at the commensurate density $n=1$ \cite{Furukawa}. 
This singularity has been interpreted as the effective mass divergence.
(2) In the insulating state, the localization length is
proportional to $\Delta^{-1/4}$, where $\Delta=\mu-\mu_{c}$ \cite{Assaad}.
Here, $\mu$ is the chemical potential and 
$\mu_{c}$ is the critical value of $\mu$ at the transition.
In terms of the scaling theory, 
these singularities lead to a new universality class
of the metal-insulator transition
with the dynamical exponent $z=1/\nu=4$
where $\nu$ is the correlation length exponent.

In strongly correlated boson systems,
the scaling theory for the superfluid-insulator transition
has also been proposed \cite{Fisher}.
Predictions from this theory have been tested favorably
by many numerical works \cite{{Batrouni},{Krauth}}.
However, bosons are usually considered as single-component objects,
compared with two spin components, up and down, in fermion systems.
The dynamical exponent $z$ is believed to be two 
at the transition between the Mott insulator and the superfluid.
Because of the absence of component degrees of freedom, 
the superfluid-insulator transition in the single-component case
is indeed the same as the transition of fermions 
from metal to the band insulator
in the sense of the universality class.

Recently, in the exact diagonalization of clusters,
the divergence of the density susceptibility
has been suggested in the two-dimensional boson $t$-$J$ model\cite{Imada2}
where the system consists of interacting two-component bosons
as we define later.
In this model, each boson has a component index $+$ or $-$  
as in the spin, up or down, of fermion systems.
Small clusters show qualitatively different behavior 
between the density and the component susceptibility
in the critical region near the Mott insulating state at $n=1$.
Especially, a remarkable result is 
a critical enhancement of the density susceptibility.
This suggests the possible existence of
a novel universality class of the superfluid-insulator transition
in the multi-component boson systems,
similarly to the metal-insulator transition in fermion systems.
The boson $t$-$J_{Z}$ model, 
which is the boson $t$-$J$ model
in the absence of the component exchange interaction, $J_{XY}=0$, 
has also been investigated \cite{Imada2}.
This model exhibits a continuous transition 
to the component-ordered state at a finite hole concentration.
Although critical exponents have not been quantitatively estimated,
the chemical potential $\mu$ in the critical region
indicates different aspects from that in the transition with $z=1/\nu=2$.
More specifically, the divergence of the density susceptibility
at the transition has been suggested.
In this previous study, the superfluidity of 
the quantum liquid state has not been explicitly investigated. 

One of the purposes of the present work is
to study the transition to the component-ordered state
in two-component boson systems in more detail.
We calculate the superfluid density 
to identify the critical point of the superfluid-insulator transition.
Through the study of the component correlation around this transition,
we identify the nature of the insulating state.
The critical properties at this transition are also
investigated in detail.

The second purpose is
to clarify possible variety of the universality class
even within the boson systems,
where the transition is between the superfluid and insulator.
Our models exhibit two types of insulating states,
one with and the other without the component order.
Comparison of critical properties near the both transitions
into these two insulating states
helps us to understand roles of component symmetry breaking 
on the superfluid-insulator transition.

The third purpose of this paper is
to study the interplay of the component order transition
at an incommensurate density, $n\neq1$, 
and the transition to the Mott insulator at $n=1$.
Our models show the transition to the component-ordered state
at a finite hole density accompanied by the strong mass enhancement. 
We propose the mechanism of this remarkable enhancement of the effective mass 
near the component order transition into Ising-type order.
This mechanism of strong mass enhancement may be 
rather different from that of the mass divergence at $n=1$.
The persistence of the component order at $n\neq1$ and 
the mass enhancement are both crucially related
with the Ising-like nature of the component order.

These studies may help us to gain more insight
from a slightly different viewpoint
on the origin and the mechanism of 
the novel universality class with $z=1/\nu=4$ at the Mott transition
seen in the two-dimensional Hubbard model.
This is a first step to discuss the universal aspects
of various types of the Mott transitions
beyond the statistics of particles.
One advantage of studying boson systems is that
they can relatively easily be treated by numerical approaches,
because, for example, the negative sign difficulty
in the quantum Monte Carlo method
does not appear in many cases.

The outline of this paper is as follows.
In \S {\ref{section model}}, our model is introduced.
Our model includes many models which have been investigated before.
We briefly review these previous studies.
Methods of our calculation are explained briefly in \S {\ref{section methods}}.
We use two methods in the present study.
One is the Gutzwiller-type variational technique, 
which is the mean field type approach,
and the other is the quantum Monte Carlo method with the world-line algorithm.
We show our results in \S {\ref{section results}}.
In contradiction to the mean field prediction,
our model exhibits a strong mass enhancement when it undergoes
the continuous transition into the component-ordered insulator
at an incommensurate density.
In \S {\ref{section discussion}}, 
comparing this with other types of superfluid-insulator transitions,
we propose a relevant mechanism for our results.
Relevance of this type of mass enhancement to other problems is discussed.
Finally, \S {\ref{section summary}} is devoted to the summary.

%% file: model.tex
\section{Model}
\label{section model}

In this work, we consider the two-component boson system
on a two-dimensional square lattice.
Each boson has a component index $s=+$ or $-$,
and both have hard-core repulsion 
which restricts the Hillbert space 
by excluding the double occupancy of any bosons at the same site.
The Hamiltonian is given by
\begin{eqnarray}
\label{btVWHamiltonian}
{\cal H} &=& \sum_{\langle ij \rangle} {\cal H}_{ij} \\
&=& \sum_{\langle ij \rangle} \left[ \ - t  \sum_{s=+,-}
\left( \tilde{a}_{i,s}^{\dagger} \tilde{a}_{j,s} + \mbox{h.c.} \right)
\right. \nonumber \\
& & \Biggl. \quad \quad \quad
+ V n_{i} n_{j} + W S_{i}^{z} S_{j}^{z} \ \Biggr] ,
\end{eqnarray}
where 
\begin{equation}
\tilde{a}_{i}^{\dagger} = 
a_{i}^{\dagger} \left( 1- a_{i}^{\dagger} a_{i} \right)
\end{equation}
with $a_{i}^{\dagger}$ ($a_{i}$) 
being a boson creation (annihilation) operator for site $i$.
Summations on $\langle ij \rangle$
are over the nearest neighbor pairs.
In (\ref{btVWHamiltonian}),
the number and spin operators are defined as
\begin{eqnarray}
n_{i} &=& \sum_{s=+,-} {a}_{i,s}^{\dagger} {a}_{i,s}, \\
S_{i}^{z} &=& \frac{1}{2} \left( 
{a}_{i,+}^{\dagger} {a}_{i,+} -
{a}_{i,-}^{\dagger} {a}_{i,-} \right) .
\end{eqnarray}

We note that any type of the diagonal interaction 
between the nearest neighbor sites
can be realized by choosing proper $V$ and $W$.
This model (\ref{btVWHamiltonian}) includes, as limiting cases,
various models already investigated before:
(a) The simplest case with $V/t=0$ and $W/t=0$ corresponds to 
the one-component boson Hubbard model at $U/t \rightarrow \infty$, 
where $U$ is the on-site interaction \cite{Onogi}.
In this case, the superfluid state is realized for $0<n<1$ and
the Mott transition takes place at $n=1$.
(b) The case with $V/t \neq 0$ and $W/t=0$ is called 
the extended boson Hubbard model at $U/t \rightarrow \infty$ 
with the nearest-neighbor repulsive interaction
\cite{Scalettar}.
This model is known to exhibit the density long-range order 
around $n=1/2$ for large values of $V/t$.
(c) The case with $V/t=0$ and $W/t \neq 0$ is the same model
as the boson $t$-$J_{Z}$ model with $J_{Z}=W$ \cite{Imada2},
where the insulating state at $n=1$ has
the component long-range order of the Ising type.
Previous study in this case \cite{Imada2} has shown that
the continuous transition to the component-ordered state occurs
at a finite hole concentration $\delta = \delta_{c}$.

From these previous results in the limiting cases,
our model (\ref{btVWHamiltonian}) is expected 
to have two types of ordered states.
One is the density-ordered state and the other is the component-ordered one.
The former should take place near the density-ordered insulator at $n=1/2$
for large values of $V/t$.
In the case with $W/t \neq 0$,
this state may have the component order 
because of the effective Ising coupling
between the next nearest neighbor sites 
derived from the second order perturbation in $t/V$.
The latter, the component-ordered state, should occur near $n=1$.
The nearest neighbor repulsion $V$ may work 
to decrease a tendency for the component order away from $n=1$.
Therefore, for large values of $V/t$, we expect that
the critical hole density $\delta_{c}$ approaches zero.


%% file: method.tex
\section{Methods}
\label{section methods}

\subsection{Gutzwiller projection technique}
\label{section Gutzwiller method}

Gutzwiller projection technique was originally devised
for the fermion Hubbard model \cite{Gutzwiller}.
This technique has been applied also to strongly correlated boson systems
\cite{Rokhsar}.
Commutation relations between the boson operators
decouple the Gutzwiller wave function into a site diagonal form
so that the energy can be exactly estimated quite easily
in contrast to the fermion case where the fermion determinant
has to be estimated by some types of approximations \cite{Vollhardt} or
statistical sampling methods \cite{Yokoyama}.
In addition, because we consider hard-core bosons,
there is no Gutzwiller variational parameter 
which controls the ratio of the double occupancy.
For the single-component system with hard-core constraint,
the Gutzwiller function is explicitly given by \cite{Rokhsar}
\begin{equation}
\label{Gutzwiller w.f.}
|\Psi \rangle = \prod_{i} |\phi_{i} \rangle ,
\end{equation}
with
\begin{equation}
|\phi_{i} \rangle = 
\left(1-n \right)^{\frac{1}{2}} |0 \rangle_{i} 
+ n^{\frac{1}{2}} |1 \rangle_{i}
\end{equation}
where $|0(1)\rangle_{i}$ represents
an unoccupied (occupied) state at site $i$.
The superfluid order parameter evaluated 
by this wave function (\ref{Gutzwiller w.f.}) is
\begin{equation}
\Delta_{s} \equiv \frac{1}{L^{2}} 
\sum_{i} \langle \Psi | a_{i}^{\dagger} | \Psi \rangle^{2}
= n \left( 1-n \right) ,
\end{equation}
therefore, this Gutzwiller state shows the superfluidity 
except for $n=0$ or $1$.

Here, we extend this technique to describe ordered states.
For the model (\ref{btVWHamiltonian}) on a bipartite lattice,
we may expect the symmetry breaking, 
such as the density order or the component order.
To allow these orderings, we extend the Gutzwiller wave function
(\ref{Gutzwiller w.f.}) in the form as
\begin{equation}
\label{2 component w.f.}
| \Psi \rangle = 
\prod_{i \in A} | \phi_{i}^{A} \rangle
\prod_{j \in B} | \phi_{j}^{B} \rangle ,
\end{equation}
where $A(B)$ represents $A(B)$-sublattice and
\begin{eqnarray}
| \phi^{A} \rangle &=& 
\left( 1-\zeta \right)^{\frac{1}{2}} | 0 \rangle
+ \lambda^{\frac{1}{2}} | + \rangle
+ \left( \zeta-\lambda \right)^{\frac{1}{2}} | - \rangle \\
| \phi^{B} \rangle &=& 
\left( 1-2n+\zeta \right)^{\frac{1}{2}} | 0 \rangle
\nonumber \\
&+& \left( n-\lambda \right)^{\frac{1}{2}} | + \rangle
+ \left( n-\zeta+\lambda \right)^{\frac{1}{2}} | - \rangle
\end{eqnarray}
with two variational parameters $\lambda$ and $\zeta$.
Here, $| + (-) \rangle$ represents an occupied state
with a $+ (-)$ component boson.

This extended Gutzwiller function (\ref{2 component w.f.})
represents various states 
according to the values of $\lambda$ and $\zeta$ as follows:
(i) When $\lambda=n/2$ and $\zeta=n$, 
(\ref{2 component w.f.}) is ascribed to 
the single component case (\ref{Gutzwiller w.f.}) 
because $|\phi^{A}\rangle = |\phi^{B}\rangle$.
Therefore, for $0<n<1$, it represents the superfluid state 
without any ordering of density or spin.
(ii) In the case of $\lambda \neq n/2$ and $\zeta=n$,
we have the component-ordered state of the Ising type,
which is characterized by 
\begin{equation}
\langle S_{i \in A}^{z} \rangle =
- \langle S_{j \in B}^{z} \rangle \neq 0, \
\langle n_{i \in A} \rangle =
\langle n_{j \in B} \rangle = n.
\end{equation}
(iii) In the case with $\lambda=\zeta/2$ and $\zeta \neq n$,
the density-ordered state appears, that is,
\begin{eqnarray}
\langle n_{i \in A} \rangle - n &=&
- \left( \langle n_{j \in B} \rangle -n \right) \neq 0, \nonumber \\
\langle S_{i \in A}^{z} \rangle &=&
\langle S_{j \in B}^{z} \rangle = 0.
\end{eqnarray}
(iv) For other values of $\lambda$ and $\zeta$, 
we have the mixed state,
which is the coexistence of the component and density order defined by
\begin{equation}
\langle S_{i \in A}^{z} \rangle \neq
\langle S_{j \in B}^{z} \rangle , \
\langle n_{i \in A} \rangle \neq
\langle n_{j \in B} \rangle.
\end{equation}
In all these states, the superfluid order parameter $\Delta_{s}$
has a finite value except for $n = 0$ or $1$.

We determine the values of $\lambda$ and $\zeta$ to minimize
the expectation value of the Hamiltonian (\ref{btVWHamiltonian}).
From this minimization, we draw the Gutzwiller phase diagram.
The mean field picture on the transitions to the ordered states
is discussed in {\ref{section GWresults}}.

\subsection{Quantum Monte Carlo technique}

To get more accurate and unbiased results 
on various physical quantities in the ground state,
we investigate by the quantum Monte Carlo (QMC) method
with the world-line algorithm.
The technique which we choose is a standard one \cite{Makivic}.
First, we rewrite the partition function
using the Suzuki-Trotter formula,
\begin{eqnarray}
Z &\equiv& e^{-\beta {\cal H}} \\
&=& \lim_{M \rightarrow \infty} \left( e^{-\Delta \tau {\cal H}} \right)^{M} \\
&=& \lim_{M \rightarrow \infty} \left[ e^{-\Delta \tau 
\left( {\cal H}^{odd}_{X} 
     + {\cal H}^{odd}_{Y} 
     + {\cal H}^{even}_{X} 
     + {\cal H}^{even}_{Y} \right)}
\right]^{M} ,
\label{checker board}
\end{eqnarray}
where $\beta=T^{-1}$ (we set $\hbar = k_{B} = 1$) is the inverse temperature
and $M = \beta / \Delta \tau$ is called the Trotter number.
In (\ref{checker board}), we make the checkerboard decomposition,
where ${\cal H}^{odd(even)}_{X(Y)}$ represents
the Hamiltonian (\ref{btVWHamiltonian}) for odd(even) bonds
in the $x(y)$ direction.
These procedures map the original quantum problem in two dimensions 
into the classical one in $(2+1)$ dimensional space
composed of plaquettes on which world lines lie.
Each Monte Carlo update is accepted by the ratio of the probability
\begin{eqnarray}
& &P \left( n_{i} \left( \tau_{l} \right), n_{j} \left( \tau_{l} \right) ;
n_{i} \left( \tau_{l+1} \right), n_{j} \left( \tau_{l+1} \right) \right) 
\nonumber \\
= \langle & & n_{i} \left( \tau_{l} \right) n_{j} \left( \tau_{l} \right) |
e^{-\Delta \tau {\cal H}_{ij}} |
n_{i} \left( \tau_{l+1} \right) n_{j} \left( \tau_{l+1} \right) \rangle
\label{probability}
\end{eqnarray}
between after and before the change in configurations on each plaquette.
Here, $\tau_{l} = l \Delta \tau (l=0,1,\cdot \cdot \cdot,\ 4M)$ is
the position in the imaginary time direction.

The calculation is done for the system size 
$L \times L$, $L = 4, 6$ and $8$ 
with $\beta=10$ which is low enough temperature
to investigate the ground-state properties 
for most of our purposes\cite{Temp}.
We take periodic boundary conditions 
in both spatial and temporal directions.
Our calculation is done in a canonical ensemble, that is, 
the density $n$ is fixed in Monte Carlo steps.
The systematic errors caused by finite $M$
are proportional to $(\Delta\tau)^2$ in the lowest order.
The extrapolation to $M \rightarrow \infty$
has been done by using several values of $M$ between $40$ and $100$
for each $L$, $n$ and $\beta$.
In the actual calculation,
typically $10000$-$20000$ sequence of configurations are discarded
for the thermalization to realize the statistical equilibrium.
We actually take $100000$-$1000000$ samples for each measurement
depending on the situation.

\begin{figure}
\hfil
\epsfile{file=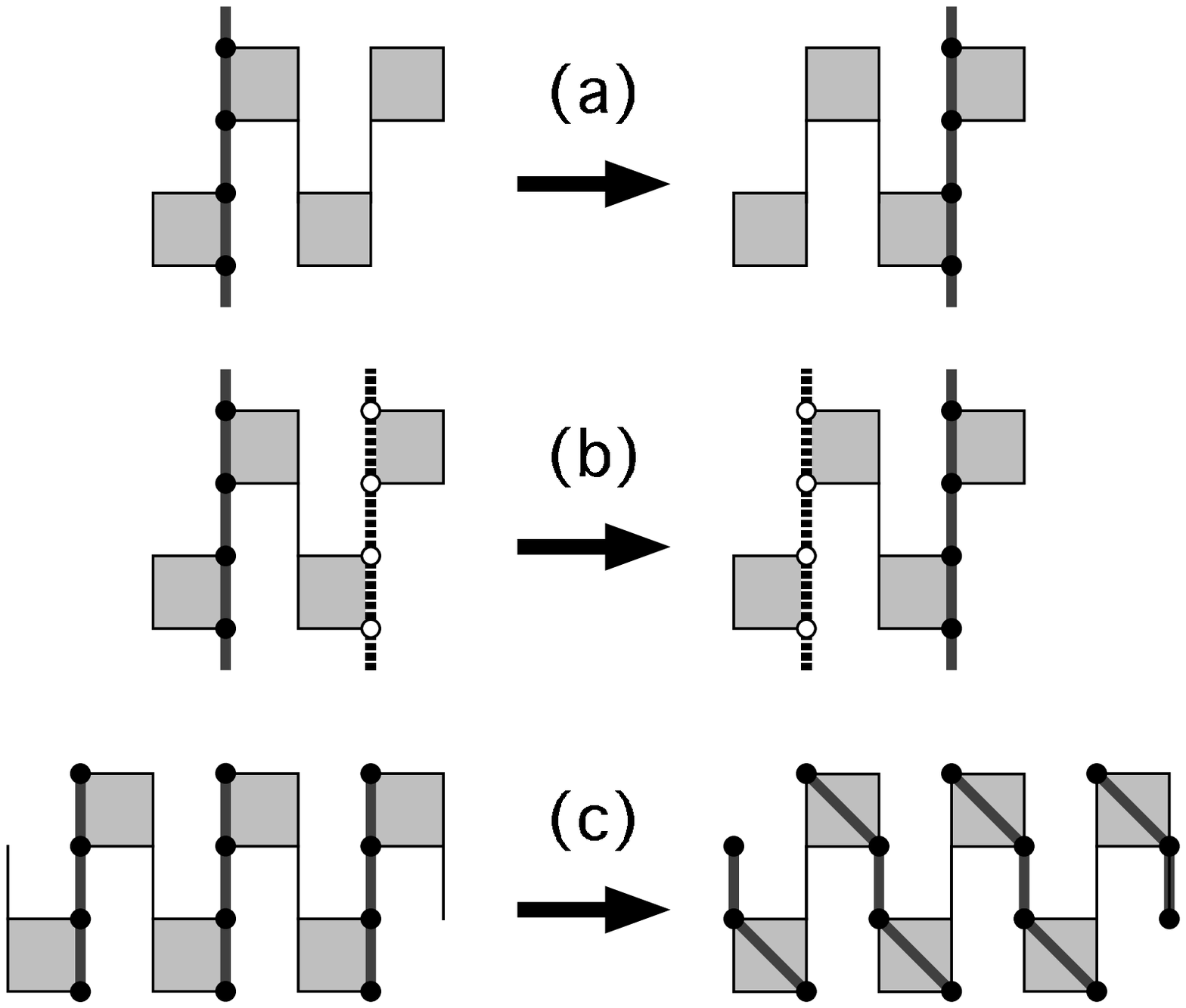,scale=0.4}
\hfil
\caption{Examples of the global flip in the QMC calculation.
In all figures, the horizontal direction is a spatial one
and the vertical direction is the temporal one.
All gray squares are plaquettes 
in the checkerboard decomposition (\ref{checker board}),
Solid and dotted lines illustrate the world lines.
Filled and open circles are bosons of the component $+$ and $-$, respectively.
(a) illustrates a uniform move of a straight world line,
(b) is for an exchange of straight world lines of different component
and (c) shows a spatial global flip which changes the winding number.}
\label{FIGflip}
\end{figure}

In the calculation,
we take two local updates and three global updates \cite{Makivic}.
The latter global updates are typically illustrated in Figure \ref{FIGflip}.
Figure \ref{FIGflip} (a) and (b) are temporal global flips, 
which are uniform moves or exchanges of straight world lines.
The latter one in Figure \ref{FIGflip} (b) can exchange 
different component bosons,
which is important near $n=1$
because of the low acceptance ratio of local updates 
caused by the growth of the component order.
The last one in Figure \ref{FIGflip} (c) is a spatial global flip
which reconnect same component world lines in every other plaquette.
This updateing procedure can change the winding number, therefore,
it is important for estimates of global physical quantities
like the superfluid density.

%% file: result.tex
\section{Results}
\label{section results}

\subsection{The Gutzwiller phase diagram and the mean field analysis
 	    on the transitions to the ordered states}
\label{section GWresults}

\begin{figure}
\hfil
\epsfile{file=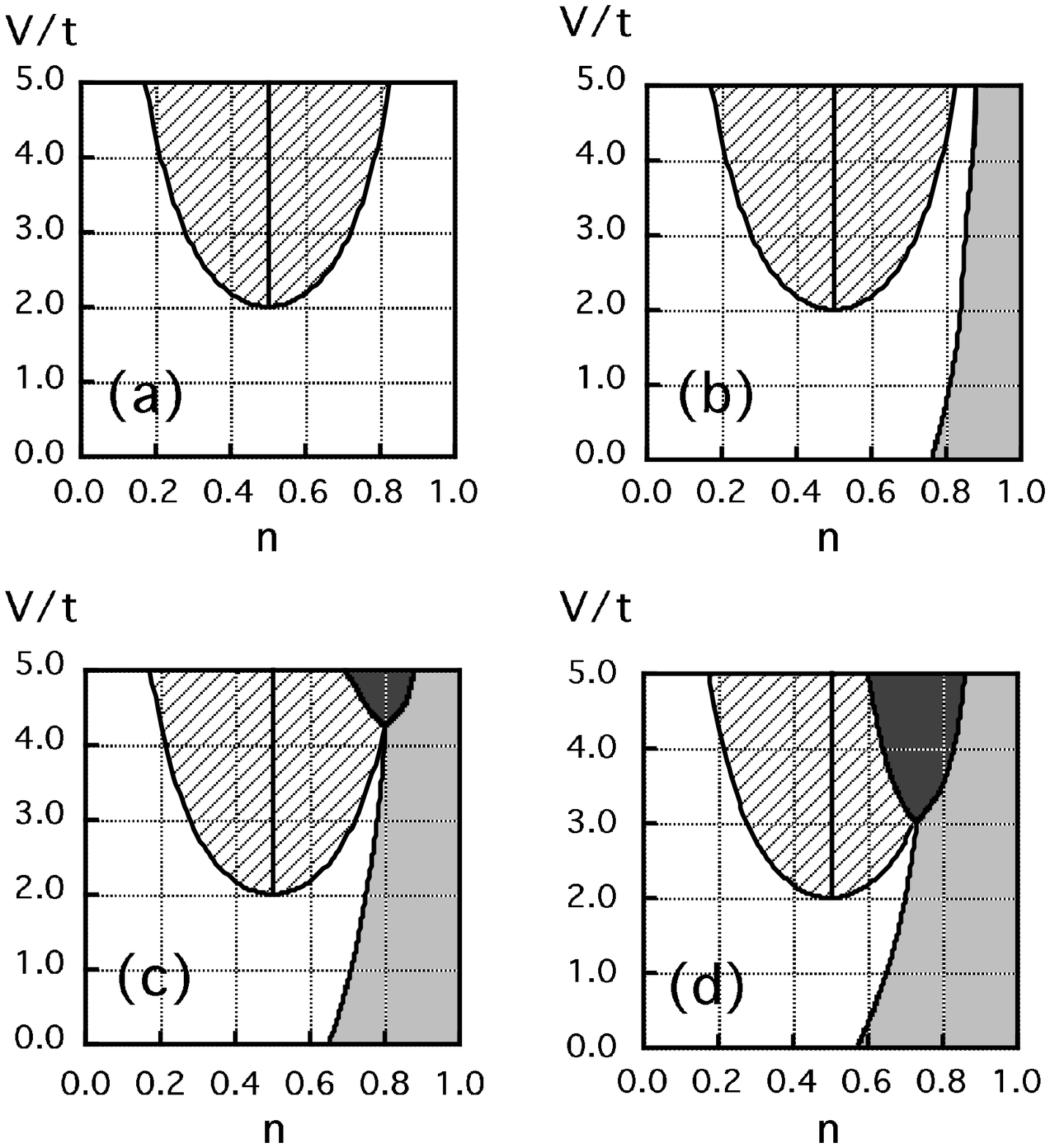,scale=0.5}
\hfil
\caption{The Gutzwiller phase diagrams.
Figures (a),(b),(c) and (d) correspond to
$W/t=0.0, 0.5, 1.0,$ and $1.5$, respectively.
Hatched areas are for the density-ordered state with superfluidity,
gray areas are the phase-separation into 
the component-ordered phase and the superfluid phase,
and black areas are for the coexistence of
the density and component order, respectively.
The straight lines at $n=1/2$ and $V/t > 2.0$ represent
the insulating states with the density order.
The superfluidity is realized in all the other regions 
except for the cases with $n=1/2$ and $V/t>2.0$,
$n=0$ and $1$.}
\label{FIGGWphase}
\end{figure}

Applying the procedure explained in \S {\ref{section Gutzwiller method}},
we determine the Gutzwiller phase diagram.
The results are depicted in Figure \ref{FIGGWphase}.

When $W/t=0$, Figure \ref{FIGGWphase} (a), 
which is equivalent to the single-component case,
shows the density-ordered state around $n=1/2$ for large values of $V/t$.
In this case, the expectation value of the Hamiltonian (\ref{btVWHamiltonian})
with the Gutzwiller-type wave function (\ref{2 component w.f.}) is given by
\begin{eqnarray}
\label{Hij W/t=0}
\langle \Psi | {\cal H}_{ij} |\Psi \rangle 
= &-& 2t \zeta^{\frac{1}{2}} \left( 2n-\zeta \right)^{\frac{1}{2}}
      \left( 1-\zeta \right)^{\frac{1}{2}}
      \left( 1-2n+\zeta \right)^{\frac{1}{2}} \nonumber \\
  &+& V \zeta \left( 2n-\zeta \right) .
\end{eqnarray}
The minimization of (\ref{Hij W/t=0}) with the variational parameter $\zeta$
gives the threshold density for the density long-range order as
\begin{equation}
n_{c} = \ \ \frac{1}{2} \left[ 1 \pm
\sqrt{ \frac{v-1}{v+1} } \ \ \right] ,
\end{equation}
where $v \equiv V/2t$.
This result agrees with the previous result 
by another type of mean field analysis 
for the extended boson Hubbard model \cite{Scalettar}.
At $n=1/2$, the insulating state with the density order appears
for $v>1.0$.
This is indicated by 
a kink of the ground state energy as the function of $n$.
In the hatched region in Figure \ref{FIGGWphase} with $n \neq 1/2$,
the superfluidity and the density order coexist in uniform phase.
These features on the density-ordered state 
persist even for $W/t\neq0$,
except for the mixed state as explained below.

When $W/t$ is switched on,
a phase-separated region appears near $n=1$
as shown in Figure \ref{FIGGWphase}.
As explained below, this is determined by a convex region 
in the curve of the ground state energy as the function of $n$.
As a simple example, we consider the $V/t=0$ case.
The expectation value of (\ref{btVWHamiltonian}) at $V/t=0$ is given by
\begin{eqnarray}
\label{Hij V/t=0}
\langle \Psi | {\cal H}_{ij} | \Psi \rangle 
= &-& 4t \left( 1-n \right) \lambda^{\frac{1}{2}} 
         \left( n-\lambda \right)^{\frac{1}{2}} \nonumber \\
  &+& \frac{W}{4} \left( 4n\lambda-4\lambda^{2}-n^{2} \right) .
\end{eqnarray}
From the minimization of (\ref{Hij V/t=0}) with $\lambda$ gives
the component-ordered state for 
\begin{equation}
\label{delta for CO}
0 \leq \delta \leq \delta^{*} = \frac{w}{w+1} ,
\end{equation}
where $w \equiv W/4t$.
However, more careful analysis shows that
this does not really happen as follows.
Because the ground state energy obtained 
by the minimization of (\ref{Hij V/t=0}) 
has a convex region adjacent to $n=1$ as the function of $n$,
a phase separation occurs into two states,
one at $n=1$ and the other at $n\neq1$.
The value of critical hole density $\delta_{c}$ 
for the phase separation line  
is determined by drawing the tangent to the ground-state energy curve
from the end point of this curve at $n=1$
in the plane of filling and energy.
In the special case with $V/t=0$ which we consider here,
this critical point $\delta_{c}$ is given as
\begin{equation}
\label{delta for PS}
\delta_{c} = \sqrt{ \frac{2w}{w+4} } .
\end{equation}
For all values of $W/t$, $\delta_{c}$ is larger than $\delta^{*}$.
Therefore, when $\delta \rightarrow 0$,
what happens in practice is
the phase separation for $0 \leq \delta \leq \delta_{c}$.
There, we have the coexistence of 
the superfluid state without the component long-range order 
at $\delta=\delta_{c}$
and the component-ordered state of the Ising type at $n=1$.
In more general, the value of $\delta_{c}$ for $V/t \neq 0$ 
is determined numerically and
$\delta_{c}$ is found to be larger than $\delta^{*}$ 
for all values of $V/t$ and $W/t$.
As shown in Figure \ref{FIGGWphase},
the larger $V/t$ gives the narrower region of this phase separation.
The mixed state defined in \S {\ref{section Gutzwiller method}}
occurs in the region beyond the intersecting point
of curves for $n_{c}$ and $\delta_{c}$, 
as shown in Figure \ref{FIGGWphase} (c) and (d).
There, large values of $V/t$ and $W/t$ cause complex orderings, however,
this mixed state is beyond our scope of the present study.

\begin{figure}
\hfil
\epsfile{file=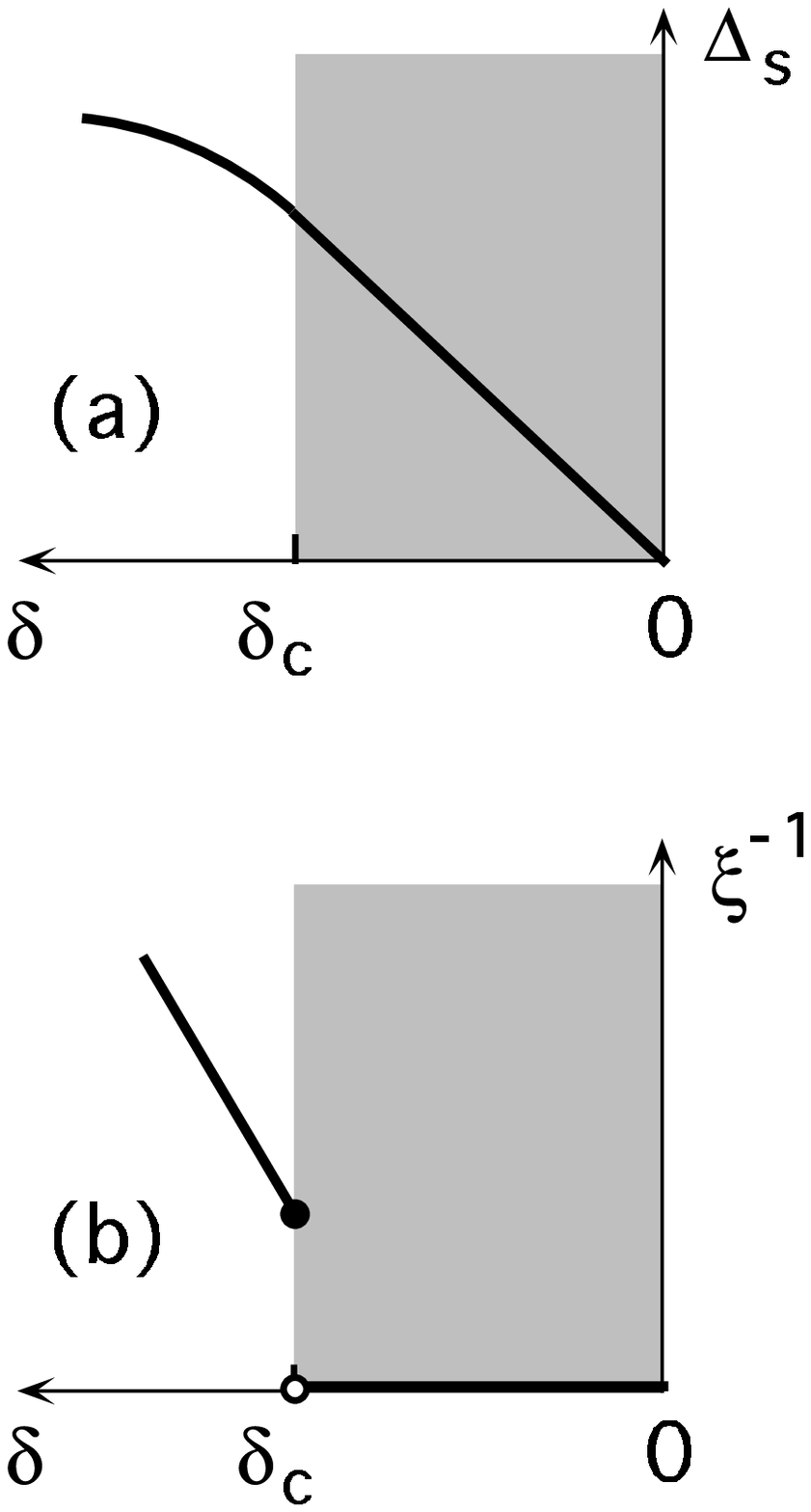,scale=0.5}
\hfil
\caption{Critical behaviors in the transition
to the phase-separated region near $n=1$
suggested from the mean field analysis.
(a) is for behavior of the superfluid order parameter and
(b) illustrates the inverse correlation length
of the component correlation.
Gray areas show the phase-separated region, as in Figure \ref{FIGGWphase}.}
\label{FIG MF prediction}
\end{figure}

These mean field analyses suggest the following pictures
of the critical properties at the transition
into the phase-separated region near $n=1$.
Here, we discuss behavior of the superfluid order parameter $\Delta_{s}$
and the correlation length $\xi$ of the component correlation.
The correlation length $\xi$ is defined from
\begin{equation}
\langle \Psi | S_{i}^{z} S_{j}^{z} | \Psi \rangle \sim
\left( - \right)^{|R_{ij}|} \exp \left( -
\frac{|\vec{r}_{ij}|}{\xi} \right) ,
\end{equation}
where $R_{ij}$ is the Manhattan distance between the site $i$ and $j$.
By definition, $\xi$ diverges in the component-ordered state.
The above Gutzwiller results suggest following behavior 
of $\Delta_{s}$ and $\xi^{-1}$, 
as schematically depicted in Figure \ref{FIG MF prediction}.
As shown in Figure \ref{FIG MF prediction} (a),
the phase separation into the superfluid state at $\delta=\delta_{c}$ 
and the insulator at $n=1$ predicts
\begin{equation}
\Delta_{s} = \frac{\delta}{\delta_{c}} 
\Delta_{s} \left( \delta=\delta_{c} \right)
\end{equation}
for $0 \leq \delta \leq \delta_{c}$.
Moreover, the absence of the component long-range order 
at $\delta=\delta_{c}$ leads to a jump of $\xi^{-1}$ 
as shown in Figure \ref{FIG MF prediction} (b).
For $0 \leq \delta < \delta_{c}$, $\xi$ diverges
because the Ising-ordered insulator at $n=1$ coexists with
the superfluid state at $\delta=\delta_{c}$ in this phase separated region.
Therefore, this mean field analysis shows that
this transition at $\delta=\delta_{c}$ is a discontinuous one
as is usual with the phase separation.

\subsection{QMC results}
\label{section QMC result}

We calculate various physical quantities by the QMC method 
for two sets of parameters.
One is the case with $V/t=2$ and $W/t=1$, and
the other is the case with $V/t=4$ and $W/t=1$.
For both cases, we change the density $n$ from $0$ to $1$.
The definitions of physical quantities are the following.
The ground state energy per site is given by
\begin{equation}
E_{\mbox{g}} = \frac{1}{L^{2}} \langle {\cal H} \rangle.
\end{equation}
The bracket defines the canonical ensemble average.
The equal-time correlation functions are defined as
\begin{eqnarray}
N \left( \vec{k} \right) &=&
\frac{1}{L^{2}} \sum_{i,j} e^{i \vec{k} \cdot \vec{r}_{ij}}
\langle n_{i} n_{j} \rangle, \\
S \left( \vec{k} \right) &=&
\frac{1}{L^{2}} \sum_{i,j} e^{i \vec{k} \cdot \vec{r}_{ij}}
\langle S_{i}^{z} S_{j}^{z} \rangle,
\label{Sk}
\end{eqnarray}
for the density and the component degrees of freedom, respectively.
The superfluid density which measures the stiffness 
under the twist of the boundary condition
is defined as \cite{Pollock},
\begin{equation}
\label{rhoS}
\rho_{s} = \frac{1}{4\beta} \langle \vec{W}^{2} \rangle,
\end{equation}
where
\begin{equation}
\vec{W} = \frac{1}{L} \sum_{i=1}^{nL^{2}} \left[ \
\vec{r}_{i} \left( \beta \right)
- \vec{r}_{i} \left( 0 \right) \ \right]
\end{equation}
is the winding number with $\vec{r}_{i} \left( \tau \right)$ being
the position of the $i$-th boson in $(2+1)$ dimensions.
A relation between the superfluid density $\rho_{s}$ and 
the superfluid order parameter $\Delta_{s}$ should be noted.
By definition, if the one-particle wave function extends over the whole system,
$\rho_{s}$ has a finite value.
On the other hand, because $\Delta_{s}$ is an order parameter 
for the superfluidity,
it has a non-zero value only in the superfluid state.
Therefore, $\rho_{s}$ and $\Delta_{s}$ may in general be different.
For example, in the free boson system in two dimensions,
we have $\rho_{s} \neq 0$ but $\Delta_{s} = 0$.
However, in the interacting boson systems,
because the ground state except for the insulating state
should be superfluid,
we may regard a state with finite $\rho_{s}$
as a superfluid state with finite $\Delta_{s}$ in the present work.

\begin{figure}
\hfil
\epsfile{file=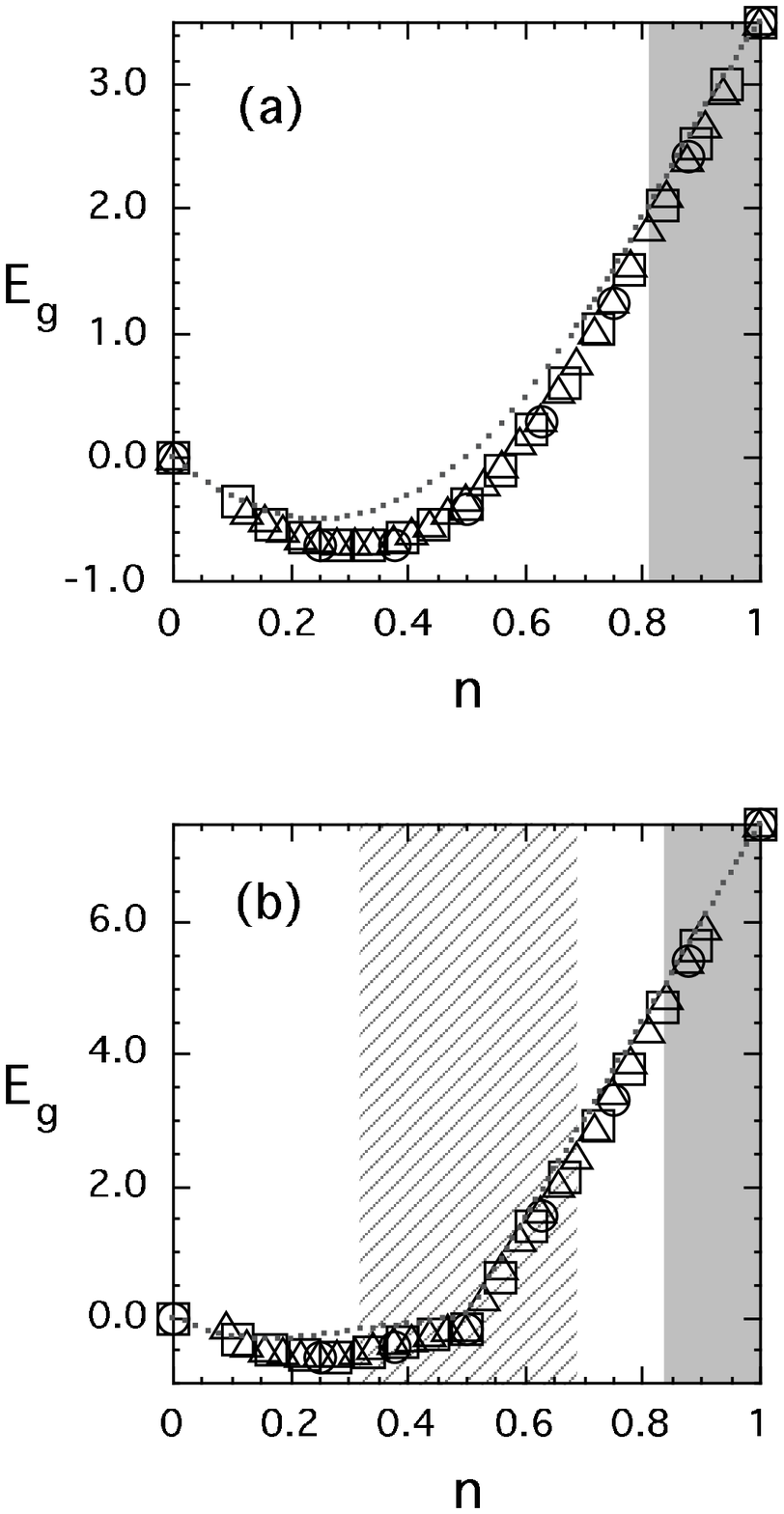,scale=0.8}
\hfil
\caption{QMC results of 
the ground state energy as the function of the density $n$.
(a) is for the case with $V/t=2$ and $W/t=1$, and
(b) is for $V/t=4$ and $W/t=1$.
Symbols are circles for $L=4$, squares for $L=6$ and triangles for $L=8$.
The gray areas represent the phase-separated regions 
determined by the convex regions of the ground-state energy curve.
The hatched area in (b) is another phase-separated region.
The dotted line in each figure is the Gutzwiller result 
obtained in \S {\ref{section GWresults}} for reference.}
\label{FIGEg}
\end{figure}

First, the ground state energy as the function of $n$
is shown in Figure \ref{FIGEg} for both cases.
Both data exhibit phase-separated regions near $n=1$, 
as in the Gutzwiller results.
These regions are narrower than those in Figure \ref{FIGGWphase}
presumably because of quantum fluctuations.
For the latter case with $V/t=4$, however,
another convex region is found around $n=1/2$.
This indicates phase separation which does not exist in the results
of the mean field level in \S {\ref{section GWresults}}.
In the QMC results, 
the curve of the ground state energy has a kink at $n=1/2$
as found in the Gutzwiller result,
which indicates that the insulating state with a finite energy gap 
is realized at $n=1/2$.
Therefore, this new phase separated region is composed of
the insulating state at $n=1/2$ and
the state at $n\neq1/2$ in the hatched area in Figure \ref{FIGEg} (b).
All the phase-separated regions in Figure \ref{FIGEg}
are determined from the fitting of the ground-state energy data
by the appropriate polynomial function of $n$.

\begin{figure}
\hfil
\epsfile{file=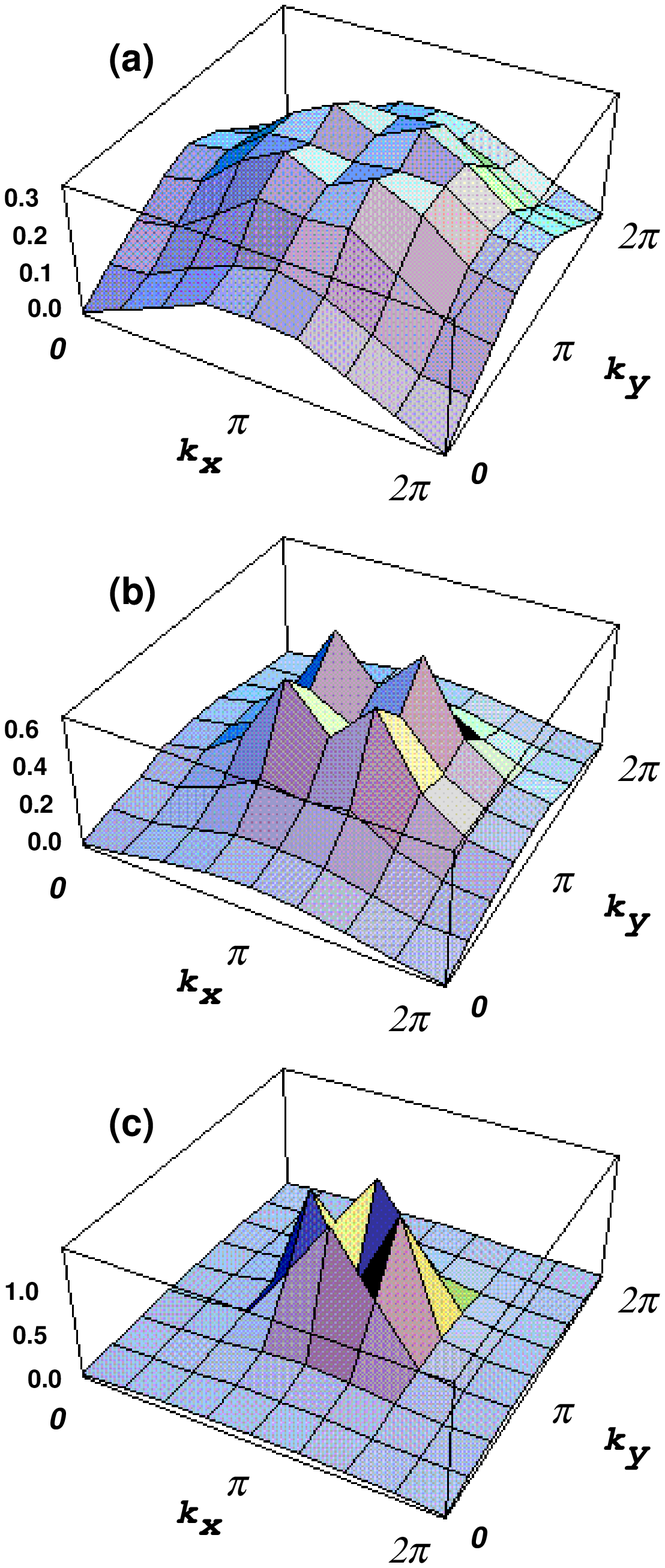,scale=0.7}
\hfil
\caption{The component correlation function (\ref{Sk})
for $V/t=4$ and $W/t=1$.
(a), (b) and (c) are for $n=46/64, 50/64$ and $54/64$.}
\label{FIGSk3Dplot}
\end{figure}

Next, other physical quantities are shown 
in Figure \ref{FIGV2W1etc} and \ref{FIGV4W1etc}.
In both cases, we find the component long-range order 
in the phase-separated regions near $n=1$ 
from Figure \ref{FIGV2W1etc} (a), (b) and \ref{FIGV4W1etc} (c), (d).
Especially, from Figure \ref{FIGV2W1etc} (b) and \ref{FIGV4W1etc} (d),
the component correlations show continuous growth
near the phase separation.
In addition, incommensurate peaks are obtained
in the component correlation function in these critical regions
for $V/t \neq 0$ as shown in Figure \ref{FIGSk3Dplot}.
In the phase-separated region,
the component correlation has commensurate peaks 
at $\vec{k} = \vec{Q} \equiv \left( \pi, \pi \right)$.

\begin{fullfigure}
\hfil
\epsfile{file=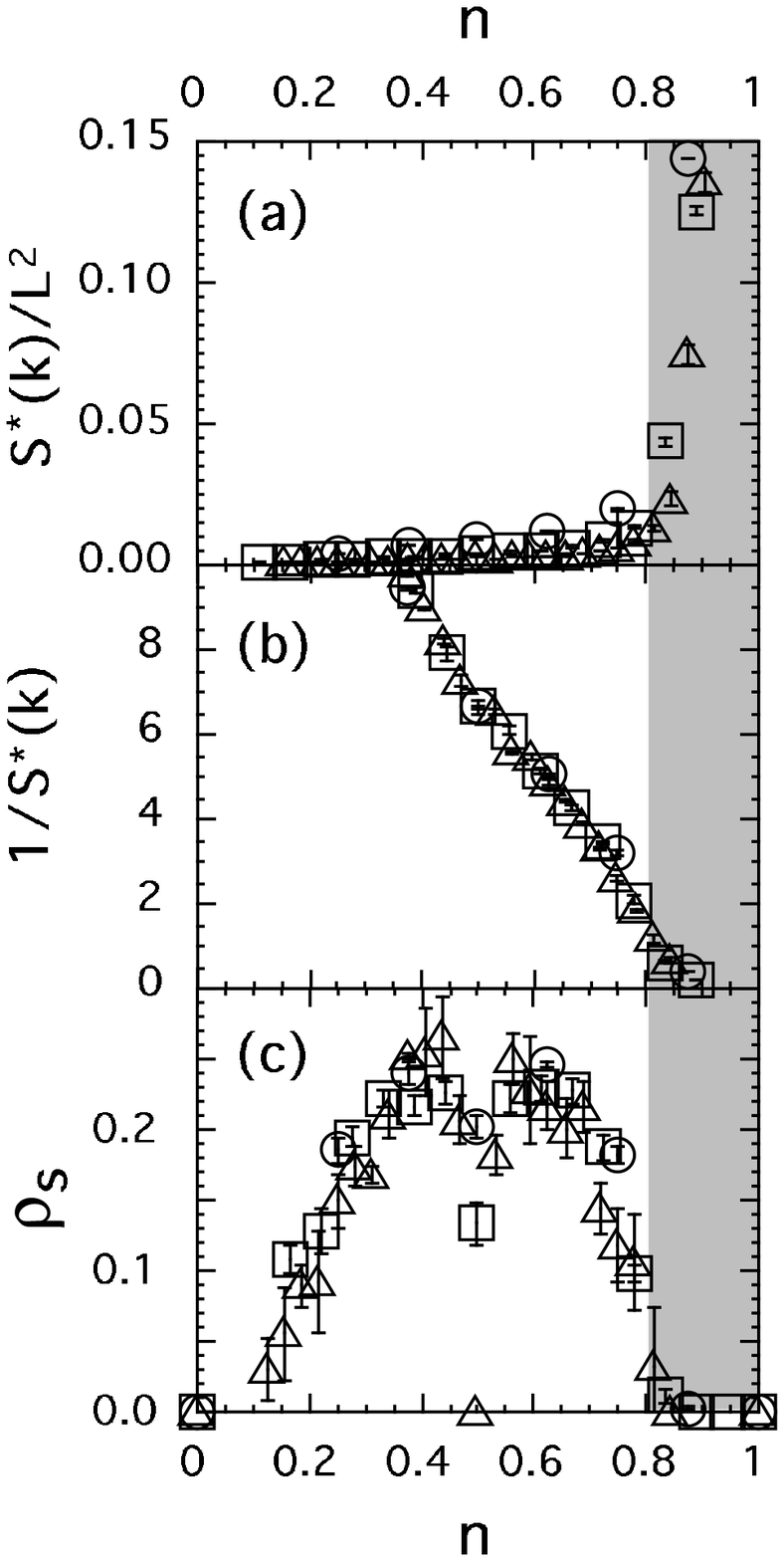,scale=0.9}
\hfil
\caption{Various physical quantities as the function of the density $n$
for the case with $V/t=2$ and $W/t=1$ obtained by the QMC calculation.
(a) shows the peak values of the component correlation function 
$S \left( \vec{k} \right)$ devided by the system size,
(b) is the plot of the inverse of the peak values of $S \left( \vec{k} \right)$
and (c) is for the superfluid density.
Symbols are circles for $L=4$, squares for $L=6$ and triangles for $L=8$.
The gray areas represent the phase-separated regions
determined by Figure \ref{FIGEg}.}
\label{FIGV2W1etc}
\end{fullfigure}

For the latter case with $V/t=4$, 
we find the density long-range order 
in the other phase-separated area around $n=1/2$
from Figure \ref{FIGV4W1etc} (a) and (b).
Similarly to the case of the component order,
the density correlation grows continuously in the critical region
as in Figure \ref{FIGV4W1etc} (b).
However, we find only the commensurate peaks 
at $\vec{k} = \vec{Q} \equiv \left( \pi, \pi \right)$
in the density correlation function for all values of the density.

It should be noted that the phase-separated region with the density order
may have a subtle problem.
When $n=1/2$ and $W/t \neq 0$, 
the second order perturbation in terms of $t/V$ gives
antiferromagnetic Ising coupling 
between next nearest neighbor pairs in the order of 
$\left(W/2 t^{2}\right) / 
\left[ \left( 3V \right)^{2} - \left( W/4 \right)^{2} \right]$.
Therefore, one might expect that some component order appears 
with the density order in the ground state at $n=1/2$ for large $V/t$
\cite{degenerate}.
Our results show no numerical evidence of this type of component order 
as in Figure \ref{FIGV4W1etc} (c).
However, the energy scale for such a component order
can be very low,
$\ll \left(W/2 t^{2}\right) / 
\left[ \left( 3V \right)^{2} - \left( W/4 \right)^{2} \right]
\sim 3.5 \times 10^{-3}$ for the present case with $V/t=4$ and $W/t=1$,
if it exists, because the frustration may suppress the ordering temperature.
All our simulations are done at $\beta=10$, which may be
too high to find this type of component order.
Therefore, the density-ordered state which appears in our result
should be considered basically 
as the ground state of the single-component system,
although whether the component order is realized or not
in the true ground state is not clear as it stands.

Though both of the ordered phases are accompanied by phase separation,
$\rho_{s}$ behaves qualitatively different in each case.
In the transition to the phase-separated region 
with the density long-range order,
$\rho_{s}$ has a finite value, as shown in Figure \ref{FIGV4W1etc} (e).
When $n \rightarrow 1/2$,
$\rho_{s}$ goes to zero continuously.
In contrast to this, in the critical region of the component order transition,
$\rho_{s}$ is strongly and continuously reduced, 
as shown in Figure \ref{FIGV2W1etc} (c) and \ref{FIGV4W1etc} (e).
In the phase-separated region,
$\rho_{s}$ is always zero within the numerical errorbars.

\begin{fullfigure}
\hfil
\epsfile{file=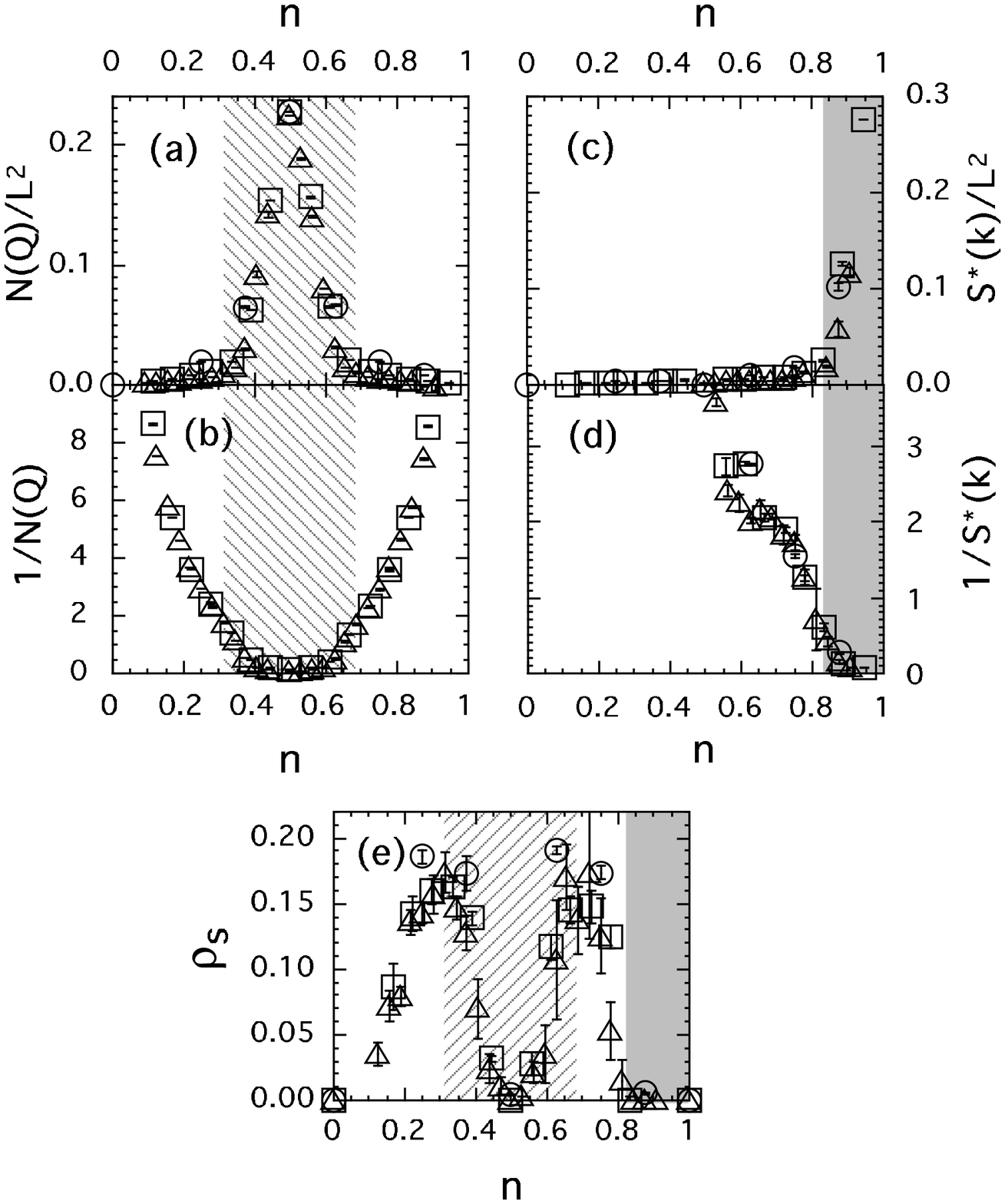,scale=0.9}
\hfil
\caption{Various physical quantities as the function of the density $n$
for the case with $V/t=4$ and $W/t=1$ obtained by the QMC calculation.
(a) illustrates the density correlation function at 
$\vec{Q} = \left( \pi, \pi \right)$, 
$N \left( \vec{Q} \right)$ devided by the system size,
(b) shows the inverse of $N \left( \vec{Q} \right)$,
(c) is for the peak values of the component correlation function 
$S \left( \vec{k} \right)$ devided by the system size,
(d) plots the inverse of the peak values of $S \left( \vec{k} \right)$
and (e) is for the superfluid density.
Symbols are circles for $L=4$, squares for $L=6$ and triangles for $L=8$.
The gray and hatched areas  represent the phase-separated regions
determined by Figure \ref{FIGEg}.}
\label{FIGV4W1etc}
\end{fullfigure}

The observed correlation functions and the superfluid density
suggest that the both transitions to the phase-separated regions 
with the density and component order are continuous
in contradiction to the results of the Gutzwiller results
in the previous subsection {\ref{section GWresults}}.

%% file: discussion.tex
\section{Discussions}
\label{section discussion}

We discuss various remarkable aspects of the QMC results 
in \S {\ref{section QMC result}}
with emphasis on the differences from the Gutzwiller results
in \S {\ref{section GWresults}}.
First, QMC results show the phase-separated region
with the density long-range order around $n=1/2$
as well as that with the component order near $n=1$
in contrast to the absence of the phase separation around $n=1/2$
in the Gutzwiller results.
The neglect of spatial fluctuations in the mean field analysis
appears to be the origin of the failure in reproducing 
the phase separation around $n=1/2$.
This phase separation may be 
a consequence of the repulsive interaction assumed only 
for on-site and the nearest neighbor pairs 
in our model (\ref{btVWHamiltonian}).
It has been suggested that the next nearest neighbor repulsion eliminates 
this type of phase separation\cite{Scalettar}.
A more important contradiction to the mean field results is that 
both transitions into the phase-separated regions with 
the component and density orders are continuous.
As seen in Figure \ref{FIGV2W1etc} (b) and \ref{FIGV4W1etc} (b), (d),
each correlation length shows continuous divergence toward the transition.
Moreover, the component correlation function, (\ref{Sk}),
shows the incommensurate peaks due to the repulsive interaction $V$.
These behaviors are clearly different from the mean field results, 
as typically shown in Figure \ref{FIG MF prediction} (b).
The third point, which is the most remarkable, is
the sharp contrast in the superfluid density
between the transitions to the density- and component-ordered state.
The persistence of $\rho_{s}$ in the phase-separated state 
with the density order suggests that
this state is composed of
the density-ordered insulator at $n=1/2$ and 
the superfluid state with the density order 
at the density of the onset of the phase separation.
This behavior of $\rho_{s}$ at the density order transition
is similar to the mean field prediction 
for the phase-separated state near $n=1$,
as shown in Figure \ref{FIG MF prediction} (a).
In contrast to this, at the onset of the component order transition,
$\rho_{s}$ vanishes within numerical errorbars, simultaneously with
the divergence of the correlation length of the Ising-type component order.
This is quite different behavior from that in the mean field analysis,
as shown in Figure \ref{FIG MF prediction} (a).
This result suggests that
the component order of the Ising type strongly suppresses the superfluidity.

\begin{figure}
\hfil
\epsfile{file=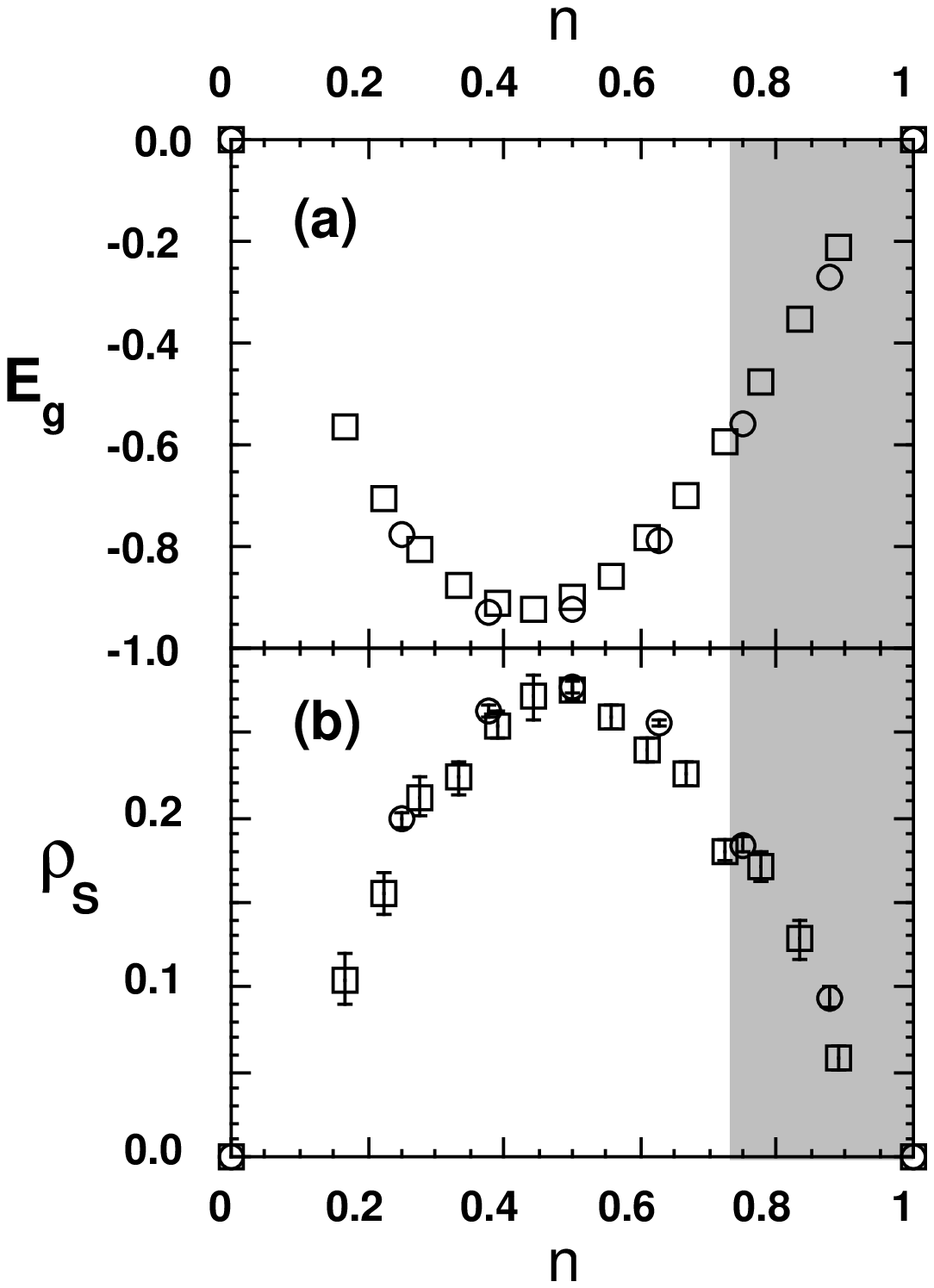,scale=0.7}
\hfil
\caption{QMC data of the ground state energy (a) and
the superfluid density (b) for the one-component model
(\ref{1compHamiltonian}) with $V/t=1$.
Symbols are circles for $L=4$ and squares for $L=6$.
The gray area represents the phase-separated region.}
\label{FIG1comp}
\end{figure}

To compare with the above results on
the remarkable suppression of $\rho_{s}$,
we consider a one-component model 
which also exhibits the phase separation near $n=1$.
Here, we take the Hamiltonian as
\begin{equation}
\label{1compHamiltonian}
{\cal H} = -t \sum_{\langle ij \rangle}
\left( \tilde{a}_{i}^{\dagger} \tilde{a}_{j} + \mbox{h.c.} \right)
+ V \left( 1-n \right) \sum_{\langle ij \rangle} n_{i}n_{j}.
\end{equation}
The QMC data of the ground state energy at $V/t=1$ 
is plotted against $n$ in Figure \ref{FIG1comp} (a).
We find that  
the nearest neighbor repulsion proportional to the hole concentration
causes a phase separation near $n=1$.
This phase-separated state is composed of
the Mott insulator at $n=1$ and 
the state at the onset of the phase separation.
The result of $\rho_{s}$ by the QMC calculation
is shown in Figure \ref{FIG1comp} (b).
We find a contrasted behavior of $\rho_{s}$
from that in the component-ordered phase of 
Figure \ref{FIGV2W1etc} (c) and \ref{FIGV4W1etc} (e).
Here, $\rho_{s}$ has a finite value in the phase-separated region and
goes to zero continuously when $n \rightarrow 1$.
This suggests that in this single-component case,
the phase separation near $n=1$ does not suppress the superfluidity.
These results support that
the suppression of the superfluidity is caused 
by the Ising-type component order,
not by the commensurability of the insulating state at $n=1$.
Phase separation in single-component systems or with the density order
is compatible with the superfluidity.

The remarkable suppression of $\rho_{s}$ in the component order transition 
in our QMC results can be interpreted as 
the strong mass enhancement as explained below.
Generally, in strongly correlated systems,
if a single-particle description is possible,
the superfluid density may be given as
\begin{equation}
\rho_{s} \propto \frac{n^{*}}{m^{*}},
\end{equation}
where $n^{*}$ is the carrier concentration and
$m^{*}$ is the effective mass in this single-particle picture.
Therefore, there are two scenarios 
to cause the superfluid-insulator transition;
$n^{*} \rightarrow 0$ or $m^{*} \rightarrow \infty$.
In fermion systems, the same classification applies,
if we consider the Drude weight instead of the superfluid density
\cite{Imada3}.
Especially, in the two-dimensional Hubbard model,
the effective mass divergence has been indicated in the Mott transition
of fermion systems \cite{{Furukawa}}.
In our results, $\rho_{s}$ is strongly reduced
at a finite hole concentration
where the component long-range order sets in.
The carrier number $n^{*}$ is apparently finite at the transition.
Therefore, the suppression of $\rho_{s}$ at $\delta=\delta_{c}$
indicates the strong enhancement of the effective mass
at the transition to the component-ordered state of the Ising type.
In contrast to this, in the cases of
the density order transition and the phase separation 
in the single-component model,
the effective mass is not enhanced near the transition.

Let us discuss the mechanism of this mass enhancement at
the transition to the component-ordered phase at $\delta=\delta_{c}$.
As shown above, this mass enhancement is caused 
by the component order of the Ising type,
not by the other orderings commonly seen even in single-component systems.
The reason why the component order survives away from $n=1$
in two dimensions against strong quantum fluctuations
appears actually to be due to the Ising-type exchange.
The mechanism of the strong mass enhancement is also ascribed to
the Ising-like nature of the component order.
As an origin of this remarkable phenomena,
we propose the hole trapping in the Ising potential, 
if the Ising order sets in.
In the fermion case, the problem of one hole in the Ising background
has been intensively studied \cite{{Kane},{Poilblanc}}.
In that case, because hopping of a hole overturns ordered spins,
the hole feels a linearly increasing potential,
so-called the string potential, with its move.
Therefore, the hole is strongly localized by this Ising potential.
If the hole hops around a unit plaquette one cycle and a half,
the hole can move to the diagonal site without disturbing Ising spins.
It may cause a weak delocalization of the hole under the Ising order.
The effective transfer to a diagonal site in a unit plaquette
in this mechanism is $32t^{6} / 675W^{5}$
in the sixth order perturbation in terms of $t/W$.
Although it is still not clear 
whether the hole is ultimately weakly delocalized 
due to the presence of this mechanism,
it is clear that the hole motion is strongly suppressed
under the Ising order and
hence the inverse effective mass and $\rho_{s}$ should be strongly reduced.
Because the transition to the Ising-ordered state in our model
happens at a finite hole density,
the analysis of the effective transfer for a single hole
doped into the Ising-ordered insulator mentioned above 
is not straightforwardly applied.
However, because the hole density is small at $\delta=\delta_{c}$,
this scenario of the mass enhancement may be essentially correct.
As it stands, it is not clear enough
whether the mass $m^{*}$ is strongly enhanced or 
really diverges at $\delta=\delta_{c}$
because of the limitation of this numerical analysis.
In any case, at least strong enhancement of $m^{*}$ and ${\rho_{s}}^{-1}$
takes place at the transition to the component-ordered state.
Furthermore, it should be noted that
the continuous growth of the Ising-type component correlation leads to 
continuous and remarkable reduction of $\rho_{s}$ and ${m^{*}}^{-1}$
even at $\delta > \delta_{c}$.
We note this mechanism of the mass enhancement is relevant
when the next nearest neighbor transfer is zero or small.
When longer range transfer than the nearest neighbor becomes large,
the mass enhancement will be reduced.

In contrast to the situation under the Ising ordering,
in the case of the transition to the density-ordered state,
bosons can move without feeling the increasing potential
like the string potential under the Ising background.
Therefore, the effective mass at this transition may not be enhanced.
This may be the reason why $\rho_{s}$ remains large and finite
in this transition.
Similarly to this, in the one-component case,
$\rho_{s}$ also remains finite because holes can move
without feeling the increasing potential at the onset of the phase separation.

The novel mechanism of the strong mass enhancement proposed above,
namely, the hole trapping under the Ising-type correlation 
even at an incommensurate density,
is likely to be effective
irrespective of the statistics of particles.
That is, if a system has Ising-type degrees of freedom 
and the Ising order survives to an incommensurate density,
the transition to the Ising-ordered state should show
the same type of continuous mass enhancement
if the range of the transfer $t$ is limited.
Here, we discuss a possible relevance of 
this type of mass enhancement in fermion systems.
In the two-dimensional single-band Hubbard model,
the component order, in more conventional word, the antiferromagnetic order
is not numerically observed away from the commensurate density $n=1$ 
presumably due to strong quantum fluctuations.
When the component (namely, spin or orbital) order is stabilized
in two-dimensional electron systems away from the Mott insulator,
it also implies that
some additional mechanism such as Ising-type anisotropy is
necessary to suppress the quantum fluctuation.
Our result suggests that
the stabilization mechanism by the Ising anisotropy
may also be the origin of strong mass enhancement
near the component order transition at an incommensurate density $n\neq1$
in two dimensions as well as in three dimensions.
Although the real phase separation is suppressed
by the long-range Coulomb interaction in electron systems,
similar mass enhancement may be observed 
for the orbital order transition because the orbital degrees 
may have anisotropic nature in the exchange process\cite{Kugel}.
It is conceivable that
this mechanism of mass enhancement is relevant
in the enhancement of the susceptibility $\chi$ and
the specific heat coefficient $\gamma$
for $\left( {\rm Y}_{1-x} {\rm Ca}_{x} \right) {\rm Ti O}_{3}$
near the metal-insulator transition point $x \sim 0.35$
which is far away from the Mott insulator at $x=0$ \cite{Kumagai}.
To judge the relevance of this proposal, 
it is desired to clarify experimentally 
whether the metal-insulator transition in 
$\left( {\rm Y}_{1-x} {\rm Ca}_{x} \right) {\rm Ti O}_{3}$
is accompanied by the orbital order transition.

The present study is a first step for a better understanding
of the variety and universal aspects of 
transitions between the quantum liquid and the insulator
beyond the statistics of particles.
Several problems are left for further study.
First, to put all the above discussions on more quantitative level,
further detailed investigation is necessary.
Especially, it is important to estimate the critical exponents
in the transition to the component-ordered state.
An interesting question is the relation of 
this transition at the incommensurate density
to the novel universality class of the metal-insulator transition 
characterized by the mass divergence at the commensurate density.
The world-line algorithm which we have used is not efficient
enough to give such quantitative details.
An interesting approach for this purpose is to employ
a model in which phase-separated region can be continuously 
reduced to the point, $\delta=0$.
This may help our further understanding of differences 
between the mass enhancement at the incommensurate density $n\neq1$ and
the commensurate density $n=1$.
In our calculations, $\delta_{c}$ decreases for the large value of $V/t$,
although the phase-separated area remains finite.
From the viewpoint of our numerical approach, 
large $V/t$ yields an additional difficulty,
that is, the acceptance ratio decreases in QMC updates,
which leads to larger errorbars.
Another possible way to reduce $\delta_{c}$ would be
to include a component-exchange term,
such as $J_{XY} \tilde{a}_{i,+}^{\dagger} \tilde{a}_{i,-}
\tilde{a}_{j,-}^{\dagger} \tilde{a}_{j,+}$.
At the present stage,
it is difficult to include this process in the Monte Carlo study
because of the sign problem.

%% file: summary.tex
\section{Summary}
\label{section summary}

In this work, we have investigated the critical properties of
various superfluid-insulator transitions in two dimensions.
We have mainly considered the two-component lattice boson system
with hard-core repulsion on a square lattice.
The nearest neighbor interaction is taken component dependent.
Our models exhibit two types of ordered states,
the density order around $n=1/2$ and the component order near $n=1$.
The Gutzwiller-type analysis shows
uniform coexistence of the superfluidity and the density order
near the density-ordered insulator at $n=1/2$ for $V/t > 2$, 
whereas the phase separation with the Ising-type component order 
in one of the phases near $n=1$.
The transition from the superfluid 
to this phase separated state with the Ising-type ordering
is a discontinuous one at the mean field level.
The QMC study indicates several remarkable properties
which are not predicted from the mean field results.
The transition to the density-ordered phase as well as
to the component-ordered phase is accompanied by phase separation.
Moreover, in contradiction to the Gutzwiller results,
all these transitions are continuous with the divergence of
the correlation length of each order.
This provides us with interesting examples of
continuous phase transition which triggers phase separation.
In spite of these similarities between the transitions to 
the density- and component-ordered phase,
QMC results show qualitative differences between them.
The most remarkable point is that
the superfluid density $\rho_{s}$ is severely reduced
in the transition to the component-ordered state,
whereas $\rho_{s} \neq 0$ in the phase-separated state
with the density order.
We have also investigated a single-component system
which has the phase separation near the Mott insulator $n=1$.
There, $\rho_{s}$ has a finite value at and around the transition.
These results suggest that
the superfluidity is suppressed by the component order of the Ising type
but not by the orderings of the single-component origin.
The second difference is seen in the correlation functions 
in each critical region of the density order transition
and the component order transition.
That is, the density correlation function exhibits only 
the commensurate peaks at $\vec{k} = \vec{Q}$ , 
while the component correlation function has the incommensurate peaks
when $V/t \neq 0$.

A remarkable consequence of our result is that
there exists a mechanism of strong mass enhancement
near the component order transition of the Ising type even when 
that transition takes place away from the commensurate density $n=1$.
The Ising exchange which presumably stabilizes
the component order away from the commensurate density $n=1$
also causes the mass enhancement.
We have proposed a picture of the hole trapping under the Ising ordering
as the origin of this type of mass enhancement.
The present mechanism of the mass enhancement is different
from the mass divergence at the commensurate density
clarified in the two-dimensional Hubbard model.
Detailed comparison between these two types of singularities 
is left for further study.
We have also discussed a possible relevance of this type of mass enhancement
to the metal-insulator transition of 
$\left( {\rm Y}_{1-x} {\rm Ca}_{x} \right) {\rm Ti O}_{3}$
at the incommensurate density $x \sim 0.35$.
In this scenario, the orbital order is stabilized 
by the Ising-type stabilization mechanism
at densities away from the filling of the Mott insulator
and this mechanism also induces the mass enhancement 
as well as the metal-insulator transition itself
if combined with finite disorder.
Detailed analysis on a quantitative level on a microscopic model 
remains for further studies.

%% file: acknowledge.tex
\section*{Acknowledgement}

This work is supported by a Grant-in-Aid for Scientific Research
on the Priority Area 'Anomalous Metallic State near the Mott Transition'
from the Ministry of Education, Science and Culture, Japan.

%% file: btVW.bbl
\begin{thebibliography}{99}

  \bibitem{Imada1}
       	M. Imada :
         J. Phys. Soc. Jpn. {\bf   64} (1995) 2956.
  \bibitem{Furukawa}
        N. Furukawa and M. Imada :
         J. Phys. Soc. Jpn. {\bf   61} (1992) 3331.
  \bibitem{Assaad}
	F. F. Assaad and M. Imada :
	 submitted to Phys. Rev. Lett.
  \bibitem{Fisher}
	M. P. A. Fisher, P. B. Weichman, G. Grinstein and D. S. Fisher :
	 Phys. Rev. B {\bf   40} (1989) 546.
  \bibitem{Batrouni}
	G. G. Batrouni, R. T. Scalettar and G. T. Zimanyi :
	 Phys. Rev. Lett. {\bf   65} (1990) 1765;
	G. G. Batrouni, R. T. Scalettar and G. T. Zimanyi :
	 Phys. Rev. B {\bf   46} (1992) 9051.
  \bibitem{Krauth}
	W. Krauth and N. Trivedi :
	 Europhys. Lett. {\bf   14} (1991) 627.
  \bibitem{Imada2}
	M. Imada :
	 J. Phys. Soc. Jpn. {\bf   63} (1994) 3059.
  \bibitem{Onogi}
	T. Onogi and Y. Murayama :
	 Phys. Rev. B {\bf   49} (1994) 9009.
  \bibitem{Scalettar}
	G. G. Batrouni, R. T. Scalettar, G. T. Zimanyi and A. P. Kampf :
	 Phys. Rev. Lett. {\bf   74} (1995) 2527;
	R. T. Scalettar, G. G. Batrouni, A. P. Kampf and G. T. Zimanyi :
	 Phys. Rev. B {\bf   51} (1995) 8467.
  \bibitem{Gutzwiller}
	M. C. Gutzwiller :
	 Phys. Rev. {\bf   134} (1964) A923.
  \bibitem{Rokhsar}
	D. S. Rokhsar and K. G. Kotliar :
	 Phys. Rev. B {\bf   44} (1991) 10328.
  \bibitem{Vollhardt}
	D. Vollhardt :
	 Rev. Mod. Phys. {\bf   56} (1985) 99 and references there in.
  \bibitem{Yokoyama}
	H. Yokoyama and H. Shiba :
	 J. Phys. Soc. Jpn. {\bf   56} (1987) 1490.
  \bibitem{Makivic}
	M. S. Makivi\'c and H. -Q. Ding :
	 Phys. Rev. B {\bf   43} (1991) 3562.
  \bibitem{Pollock}
	E. L. Pollock and D. M. Ceperley :
	 Phys. Rev. B {\bf   36} (1987) 8343.
  \bibitem{Temp}
	This temperature is not low enough to investigate the component order
	in the density-ordered state. We comment this point
	in \S {\ref{section QMC result}}.
  \bibitem{degenerate}
	In the classical consideration, however,
	the ground state has macroscopic degrees of degeneracy 
 	due to strong frustration of the nearest neighbor
	and the next nearest neighbor couplings.
  \bibitem{Imada3}
	M. Imada :
	 J. Phys. Soc. Jpn. {\bf   62} (1993) 1105.
  \bibitem{Kane}
	C. L. Kane, P. A. Lee and N. Read :
	 Phys. Rev. B {\bf   39} (1989) 6880.
  \bibitem{Poilblanc}
	D. Poilblanc, T. Ziman and E. Dagotto :
	 Phys. Rev. B {\bf   47} (1993) 14267.
  \bibitem{Kugel}
	K.I.Kugel' and D.I.Khomskii :
	 Usp.Fiz.Nauk. {\bf   136} (1982) 621,
	 [Sov.Phys.Usp. {\bf   25} (1982) 231] and references there in.
  \bibitem{Kumagai}
	K. Kumagai, T Suzuki, Y. Taguchi, Y. Okada,
	Y. Fujishima and Y. Tokura :
	 Phys. Rev. B {\bf   48} (1993) 7636.
\end{thebibliography}
